# Algorithmic Cooling of Nuclear Spin Pairs using a Long-Lived Singlet State

## Short title: Algorithmic Cooling of Nuclear Spins

Bogdan A. Rodin,[1] Christian Bengs,[2] Lynda J. Brown,[2] Kirill F. Sheberstov,[3] Alexey S. Kiryutin,[1,4] Richard C. D. Brown,[2] Alexandra V. Yurkovskaya[1,4], Konstantin L. Ivanov,[1,4] and Malcolm H. Levitt[2*]

[1] International Tomography Center SB RAS, Novosibirsk, Russia

[2] Department of Chemistry, Southampton University, Southampton SO17 1BJ, UK

[3] Johannes Gutenberg-Universität, Helmholtz Institute Mainz, Mainz 55099, Germany

[4] Novosibirsk State University, Novosibirsk, Russia

**Abstract**

Algorithmic cooling methods manipulate an open quantum system in order to lower its temperature below that of the environment. We show that significant cooling is achieved on an ensemble of spin-pair systems by exploiting the long-lived nuclear singlet state, which is an antisymmetric quantum superposition of the "up" and "down" qubit states. The effect is demonstrated by nuclear magnetic resonance (NMR) experiments on a molecular system containing a coupled pair of near-equivalent $^{13}C$ nuclei. The populations of the system are subjected to a repeating sequence of cyclic permutations separated by relaxation intervals. The long-lived nuclear singlet order is pumped well beyond the unitary limit, and the nuclear magnetization is enhanced by 21% relative to its thermal equilibrium value. To our knowledge this is the first demonstration of algorithmic cooling using a quantum superposition state and without making a distinction between rapidly and slowly relaxing qubits.

[145 words]

**MAIN TEXT**

**Introduction**

A quantum system which is in contact with a thermal reservoir, and left undisturbed, eventually comes into equilibrium, so that the populations of the quantum states are described by the Boltzmann distribution at the reservoir temperature. In the case of a nuclear spin system in a strong magnetic field, this leads to a small thermal equilibrium magnetization parallel to the field. In most nuclear magnetic resonance (NMR) experiments, the thermal nuclear magnetization is subjected to a sequence of coherent manipulations, eventually generating transverse magnetization perpendicular to the field. The precession of the transverse magnetization around the field induces the NMR signal through Faraday induction.

In the majority of NMR experiments the initial spin order established by thermal equilibration is transformed into different spin order modes, with the total quantity of spin order being conserved (in the best case), or decreased by irreversible decoherence and relaxation. Nevertheless, there are some situations in which the on-going contact of the spin system and the thermal molecular environment may be exploited to increase the amount of nuclear spin order, at least locally. A seminal example is the steady-state nuclear Overhauser effect, in which radiofrequency irradiation which is resonant with one nuclear spin species leads, under certain conditions, to an increase in the magnetization of a second spin species (*1*). This class of phenomenon has become known as *heat-bath algorithmic cooling* and has received much theoretical and experimental attention (*2–11*). Algorithmic cooling methods achieve, in favourable cases, an increase in nuclear polarization beyond its thermal equilibrium value, and hence a decrease in spin temperature. NMR algorithmic cooling evolved from the field of NMR quantum computation (*12*), and is based on a view of individual nuclear spins as independent qubits which may have distinct environmental couplings and which are addressable by frequency-selective radiofrequency fields.

Heat-bath algorithmic cooling has been demonstrated on a spin system containing two different nuclear spin species, one of which relaxes rapidly (and is therefore strongly coupled to the thermal environment) while the second species relaxes slowly (and therefore has a much weaker environmental coupling) (*2–7*). The magnetization of the slowly relaxing species is used to temporarily store the initial spin order, while the rapidly relaxing species acquires a new batch of spin order through contact with the thermal environment. The total entropy of the nuclear spin system is thereby decreased below its initial value, allowing an increase in the total spin magnetization after further manipulations. However, this particular scheme is restricted to spin systems with two species having very different relaxation properties, and with sufficiently well-resolved spectra that each species may be addressed individually by controlling the frequencies of the radiofrequency pulses. These are very restrictive conditions.

In this paper we demonstrate the algorithmic cooling of an ensemble of spin-1/2 pair systems which exhibit near-magnetic-equivalence (*13–17*), meaning that a small difference in chemical environments induces a difference in Larmor frequencies which is much smaller than the scalar spin-spin coupling. This appears to be an unpromising substrate for algorithmic cooling for the following reasons: (i) the individual spins-1/2 (qubits) are tightly coupled, do not display resolved spectral resonances, and cannot be addressed selectively using resonant irradiation; (ii) two of the four energy eigenstates are entangled superpositions of the "up-down" and "down-up" qubit states; (iii) the dominant relaxation mechanisms, which couple the quantum system to the environment, are the dipole-dipole (DD) coupling and chemical shift anisotropy (CSA) mechanisms, which affect both spins equally. Unlike previous algorithmic cooling demonstrations, no distinction may be made between fast-relaxing and slowly relaxing qubits.

We implement heat-bath algorithmic cooling of this tightly coupled spin-1/2 pair system by exploiting the special properties of the *nuclear singlet state* (*13–32*), which is an antisymmetric entangled state of the two spins-1/2, as follows:

$$|1\rangle = \frac{1}{\sqrt{2}}\left(|\alpha\beta\rangle - |\beta\alpha\rangle\right) \qquad (1)$$

In many cases, including the molecular system under study here, the nuclear singlet state $|1\rangle$ is very weakly coupled to the environment under the dominant relaxation mechanisms, namely the DD coupling and the correlated part of the CSA mechanism (*15, 16, 20*). In contrast, the three components of the triplet state, defined as follows:

$$\begin{aligned}|2\rangle &= |\alpha\alpha\rangle \\ |3\rangle &= \frac{1}{\sqrt{2}}\left(|\alpha\beta\rangle + |\beta\alpha\rangle\right) \\ |4\rangle &= |\beta\beta\rangle\end{aligned} \qquad (2)$$

are relatively strongly coupled with the environment through both the DD and the CSA mechanisms. The weak environmental coupling of the singlet state is due to strong correlations in the fluctuating fields and has been exploited in numerous NMR experiments (*9, 13–31*). In this work we show that the singlet state may be used as a temporary storage register in an algorithmic cooling protocol. The singlet state may be used as a protected repository of accumulated spin order, allowing a fresh batch of spin order to be accumulated by thermal polarization of the triplet manifold. This decreases the overall entropy of the spin system, beyond the Shannon bound (*2*). Further manipulations generate nuclear spin magnetization which is significantly enhanced with respect to its thermal equilibrium value, corresponding to a reduction in the effective spin temperature.

**Results**

*Cooling protocol.* Thermal equilibration of the spin-1/2 pair system with the molecular environment generates a set of populations given by

$$\mathbf{p}_{eq} = \begin{pmatrix} p_1^{eq} \\ p_2^{eq} \\ p_3^{eq} \\ p_4^{eq} \end{pmatrix} = \frac{1}{4} \begin{pmatrix} 1 \\ 1+\varepsilon \\ 1 \\ 1-\varepsilon \end{pmatrix} \tag{3}$$

where $\varepsilon = \hbar \gamma B^0 / k_B T$, the magnetogyric ratio of the spins is $\gamma$, the external magnetic field is $B^0$, and the environmental temperature is $T$. The states are numbered as in equations (1) and (2).

The Zeeman and singlet orders of the spin system are defined as follows:

$$\begin{aligned} \langle ZO \rangle &= N_{ZO}(p_2 - p_4) \\ \langle SO \rangle &= N_{SO}\left(p_1 - \frac{1}{3}(p_2 + p_3 + p_4)\right) \end{aligned} \tag{4}$$

where the population of a state $|s\rangle$ is denoted $p_s$ and the normalization factors are $N_{ZO} = 1/\sqrt{2}$ and $N_{SO} = \sqrt{3}/2$. The Zeeman order in thermal equilibrium is given by

$$\langle ZO \rangle_{eq} = \frac{\varepsilon}{2\sqrt{2}} \tag{5}$$

There is no singlet order in thermal equilibrium ($\langle SO \rangle_{eq} = 0$). Several procedures exist for the coherent conversion of Zeeman order into singlet order (*13, 14, 17–19, 21, 22, 24, 26, 28, 30*). All such procedures are limited by the conservation of entropy (Shannon bound (*2, 3*)), and even more restrictively, by the Sørensen bound $|\langle SO \rangle| \leq \langle SO \rangle_U^{max}$, where

$$\langle SO \rangle_U^{max} = \sqrt{\frac{2}{3}} \langle ZO \rangle_{eq} \tag{6}$$

This bound may be derived from the eigenvalue spectra of the relevant spin operators (*33–35*).

Several algorithmic cooling protocols are feasible. The protocol discussed here is composed of two classes of spin system manipulation:

(i) *Cyclic permutations* of the spin state populations, denoted $\pi_{abc}$, and described by the following transformations:

$$\begin{aligned}\pi_{abc}|a\rangle &= |b\rangle e^{i\varphi} \\ \pi_{abc}|b\rangle &= |c\rangle e^{i\varphi'} \\ \pi_{abc}|c\rangle &= |a\rangle e^{i\varphi''}\end{aligned} \tag{7}$$

where the symbols $\varphi$, $\varphi'$ and $\varphi''$ denote arbitrary phase factors. The cooling protocol exploits cyclic permutations of the populations of the three states $\{|1\rangle, |2\rangle, |4\rangle\}$ in opposite senses, described by the following matrices:

$$\pi_{124} = \begin{pmatrix} 0 & 0 & 0 & 1 \\ 1 & 0 & 0 & 0 \\ 0 & 0 & 1 & 0 \\ 0 & 1 & 0 & 0 \end{pmatrix} ; \quad \pi_{142} = \begin{pmatrix} 0 & 1 & 0 & 0 \\ 0 & 0 & 0 & 1 \\ 0 & 0 & 1 & 0 \\ 1 & 0 & 0 & 0 \end{pmatrix} \tag{8}$$

(ii) The *triplet thermal reset*, denoted $\Theta_T^{reset}$, which brings the three triplet populations into thermal equilibrium with the molecular environment, while leaving the singlet population unchanged. In the high-temperature limit, the triplet thermal reset is described by the following matrix:

$$\Theta_T^{reset} = \begin{pmatrix} 1 & 0 & 0 & 0 \\ 0 & \frac{1}{3}(1+\varepsilon) & \frac{1}{3}(1+\varepsilon) & \frac{1}{3}(1+\varepsilon) \\ 0 & \frac{1}{3} & \frac{1}{3} & \frac{1}{3} \\ 0 & \frac{1}{3}(1-\varepsilon) & \frac{1}{3}(1-\varepsilon) & \frac{1}{3}(1-\varepsilon) \end{pmatrix} \tag{9}$$

The thermal equilibrium population vector in equation (3) is an eigenvector of this matrix, to first order in $\varepsilon$, with eigenvalue 1.

The cooling protocol involves the application of $n_P$ permutation operations, alternating with triplet thermal resets. The protocol depends on whether $n_P$ is even or odd, as follows:

$$S(n_P) = \begin{cases} C^{n_P/2} & \text{for even } n_P \\ \left(C^{(n_P-1)/2}, \Theta_T^{reset}, \pi_{124}\right) & \text{for odd } n_P \end{cases} \tag{10}$$

where $C^n$ denotes $n$ repetitions of the cycle $C$, given by

$$C = \left( \Theta_T^{reset}, \pi_{124}, \Theta_T^{reset}, \pi_{142} \right) \tag{11}$$

These sequences are read from left to right in chronological order.

If a sequence $S(n_P)$ is applied to a system in thermal equilibrium, the resulting population vector may be generated by matrix-vector multiplication. In the case of an even permutation number $n_P$, this leads to

$$\mathbf{p}(n_P) = \mathbf{C}^{n_P/2} \mathbf{p}_{eq} \tag{12}$$

where the matrix representation of the cycle $C$ is given by:

$$\mathbf{C} = \pi_{142} \cdot \Theta_T^{reset} \cdot \pi_{124} \cdot \Theta_T^{reset} \tag{13}$$

In the case of an odd permutation number $n_P$, we get:

$$\mathbf{p}(n_P) = \pi_{124} \cdot \Theta_T^{reset} \cdot \mathbf{p}(n_P - 1) \tag{14}$$

In both cases, the singlet order after $n_P$ permutations is found to be

$$\langle SO \rangle (n_P) = (-1)^{n_P} \frac{\varepsilon \sqrt{3}}{4} \left( 1 - 3^{-n_P} \right) \tag{15}$$

For large values of $n_P$, this tends to the steady-state value

$$\langle SO \rangle_{ss} = (-1)^{n_P} \frac{\varepsilon \sqrt{3}}{4} = (-1)^{n_P} \sqrt{\frac{3}{2}} \langle ZO \rangle_{eq} = (-1)^{n_P} \frac{3}{2} \langle SO \rangle_U^{max} \tag{16}$$

which alternates in sign for odd and even $n_P$.

In the steady-state, and in ideal circumstances, the magnitude of singlet order is therefore enhanced by a factor of 3/2 relative to the unitary bounds on transformation of the thermal equilibrium state (equation (6)).

*Molecular System.* Practical implementation of this protocol requires a molecular system containing spin-1/2 pairs which are nearly magnetic equivalent and with a large ratio of relaxation times $T_S/T_1$. The time constant $T_S$ characterizes the decay of singlet spin order while the relaxation time $T_1$ characterizes the thermal equilibration of magnetization within the triplet manifold. A large ratio $T_S/T_1$ indicates that the triplet states are in good thermal contact with the environment, while the singlet state is only weakly coupled to the environment. This allows convenient implementation of the triplet thermal reset operation $\Theta_T^{reset}$, by allowing the system to relax for a time interval long compared to $T_1$, but short compared to $T_S$.

A suitable molecular system is given by the $^{13}C_2$-labelled naphthalene derivative shown in Figure 1(a) and referred to hereafter as $^{13}C_2$-**I**. This system has been shown to support long-lived $^{13}C_2$ singlet order with a decay time constant $T_S$ exceeding 1 hour in low magnetic field (*16*). The experiments described below were performed at a high magnetic field of 16.4 T on a sample of $^{13}C_2$ dissolved in acetone-$d_6$. Under these conditions, the relaxation time constants for the $^{13}C_2$ pair were determined to be $T_1 = 7.36 \pm 0.02$ s and $T_S = 214 \pm 5$ s respectively, corresponding to a relaxation time ratio of $T_S/T_1 = 29.1 \pm 0.7$.

The $^{13}C$ NMR spectrum of $^{13}C_2$-**I** in acetone-$d_6$ at a field of 16.4 T is shown in Figure 1(b). The spin system is well-described by a $^{13}C$-$^{13}C$ scalar coupling $J = 54.141 \pm 0.001$ Hz and a small chemical shift difference $\Delta\delta = 0.057 \pm 0.001$ ppm, due to the deliberate chemical asymmetry of $^{13}C_2$-**I**. The difference in the chemically shifted Larmor frequencies for the two $^{13}C$ sites in the 16.4 T magnetic field is $\omega_\Delta = \omega^0 \Delta\delta = 2\pi \times 10.01$ Hz. Since $|2\pi J| \ll |\omega_\Delta|$, the $^{13}C_2$ pair comprises a near-equivalent AB spin system giving rise to a four-peak pattern with two strong inner peaks and two weak outer peaks, as shown in Figure 1(b). The splitting between the strong inner peaks is given by $\sim |\omega_\Delta^2 / 8\pi^2 J| \simeq 0.9$ Hz. The outer peaks are very weak but may be enhanced by specialised pulse techniques (*28*).

***Pumping of Singlet Order.*** The build-up of singlet order under algorithmic cooling was assessed by the protocols sketched in figure 2(a,b). The intervals $\tau$ are chosen such that $T_1 \ll \tau \ll T_S$. Under these circumstances, each relaxation interval $\tau$ provides a good approximation to the triplet thermal reset $\Theta_T^{reset}$. In practice the value $\tau = 28.0$ s was used.

In all cases, a singlet-order-destruction (SOD) sequence was applied after signal detection, in order to allow faster repetition of the experiments (*32*) (not shown in figure 2).

For even permutation numbers $n_P$ (figure 2(a)), the pumping cycle *C* in equation (11) was repeated $n_P/2$ times, followed by an optional evolution interval $\tau_{ev}$, a $T_{00}$ signal filtration element (*17*, *29*), and a shaped APSOC pulse (Adiabatic Passage Spin-Order Conversion) (*24–27*). The $T_{00}$ signal filtration element suppresses all signal components that do not pass through singlet order, while the APSOC pulse efficiently converts the singlet order to transverse magnetization, which induces an NMR signal in the subsequent interval of free precession. The integrated amplitude of the NMR signal is therefore a reliable measure of the singlet order generated by the repeated cycles *C*.

For odd permutation numbers $n_P$, an extra $\tau$ delay and permutation element was inserted, and the sense of the final APSOC pulse reversed (figure 2(b)).

Fig.3 shows the singlet-filtered signal amplitude, normalized with respect to the thermal equilibrium signal induced by a 90° pulse, as a function of the number of permutations $n_P$, using a triplet thermal reset interval $\tau = 28.0$ s, and zero evolution interval, $\tau_{ev} = 0$. As predicted qualitatively by equation (15), the singlet-filtered signal amplitude builds up rapidly, reaching a steady-state of around 0.85 times the thermal equilibrium signal after approximately 6 permutations. Further permutations do not increase the observed singlet order.

The unitary bound on the conversion of thermal equilibrium magnetization into singlet order is given by the factor $\sqrt{2/3}$, as shown in equation (6). Conversion of singlet order into observable magnetization also encounters the same bound of $\sqrt{2/3}$. Hence, the overall efficiency for passing magnetization through singlet order using coherent processes, and back again, is bounded by the factor of $\left(\sqrt{2/3}\right)^2$, which is $2/3$. This two-way magnetization-to-singlet bound is shown by the red dashed line in Fig.3. The results show that the algorithmic cooling cycles successfully generated singlet spin order with an amplitude 1.27 times the unitary limit. The loss with respect to the ideal factor of 3/2 (equation (16)) may be attributed to non-ideality of the triplet thermal reset, due to the finite ratio of $T_S$ to $T_1$, and experimental infidelities in the population permutations.

The singlet pumping efficiency depends on the delays $\tau$ for the triplet thermal reset, as shown in Figure 4(a). If the delays are too short, there is not enough time to establish fresh polarization in the triplet manifold before the next permutation; if the delays are too long, singlet-triplet transitions allow the loss of singlet order during the delays. In the case considered, the generation of singlet order is optimised for delays of duration $\tau = 28.0$ s.

The generation of pumped singlet order was verified by allowing an evolution time $\tau_{ev}$ for the singlet order to decay, before it is converted into magnetization for detection. As shown in Figure 4(b), the singlet order decays with the characteristically long time constant of $T_S = 209 \pm 6$ s, which greatly exceeds the spin-lattice relaxation time constant of $T_1 = 7.36 \pm 0.02$ s. This behaviour verifies that the state generated by the pumping protocols does indeed correspond to long-lived nuclear singlet order.

*Enhancement of Magnetization.* The low-entropy state achieved by pumping of algorithmic cooling of singlet order may be used to enhance the nuclear magnetization. In the case of even permutation numbers $n_P$, enhanced Zeeman order may be generated from the steady state by application of a further triplet thermal reset followed by a population swap between states $|1\rangle$ and $|2\rangle$, as represented by the following matrix:

$$\pi_{12} = \begin{pmatrix} 0 & 1 & 0 & 0 \\ 1 & 0 & 0 & 0 \\ 0 & 0 & 1 & 0 \\ 0 & 0 & 0 & 1 \end{pmatrix} \qquad (17)$$

The final population swap leads to a population vector $\mathbf{p}_{final} = \pi_{12} \cdot \Theta_T^{reset} \cdot \mathbf{p}_{ss}$. In the ideal case, the final Zeeman order is enhanced in magnitude by a factor of 3/2 relative to its thermal equilibrium value:

$$\langle ZO \rangle_{final} = \frac{3\varepsilon}{4\sqrt{2}} = \frac{3}{2} \langle ZO \rangle_{eq} \qquad (18)$$

This value of the Zeeman order corresponds to a nuclear spin system which has been cooled to a temperature equal to 2/3 of the environmental temperature.

Figure 2 (c) shows the experimental protocol for detecting the enhanced magnetization, in the case of even permutation numbers $n_P$. The final triplet thermal reset is implemented by an interval of duration $\tau'$, while the $\pi_{12}$ population swap is implemented by an APSOC(+) pulse. The optimal value of $\tau'$ was found empirically to be 18.0 s, which is smaller than the optimal pumping intervals $\tau = 28.0$ s.

Figure 5 shows experimental results demonstrating the enhancement of the nuclear spin magnetization with respect to its thermal equilibrium value. Figure 5(b) shows a $^{13}$C NMR signal of $^{13}$C$_2$-**I**, enhanced by 21% with respect to the thermal equilibrium spectrum in figure 5(a). This corresponds to a reduction of spin temperature by a factor of 0.83. This result proves that it is possible to enhance the nuclear magnetization by algorithmic cooling, even on a strongly coupled spin system which lacks any distinction between fast-relaxing and slowly relaxing spins.

**Discussion**

Since the achieved enhancement of nuclear magnetization is modest, we do not claim that this method is in serious competition with true hyperpolarization techniques such as dynamic nuclear polarization (*36*). Nevertheless the phenomenon is of interest, since it transcends several paradigms which have played a prominent role in the theory and implementation of algorithmic cooling. The algorithmic cooling of nuclear spin systems has generally been framed in the language of NMR quantum computation, in which a prevailing assumption is that the individual nuclear spins may be treated, to a good approximation, as independent qubits which are independently addressable by suitable radiofrequency fields (*2–7*). The current work has shown that (i) algorithmic cooling may be achieved on systems which lack independently addressable qubits; (ii) algorithmic cooling may exploit entangled superpositions of the qubit states (the singlet state in the current case); (iii) algorithmic cooling does not necessarily require qubits with distinct relaxation behaviour; (iv) algorithmic cooling may be used to enhance modes of spin order which do not correspond to spin polarization.

The last point is salient: The primary target of the cooling algorithm in the current work is not the establishment of a low effective spin temperature in the spin ensemble but the enhancement of non-magnetic singlet order. This is an anticorrelated state of the spin ensemble which does not correspond to macroscopic spin polarization and which cannot be described by a spin temperature. Nevertheless, as shown here, singlet order may be pumped by an algorithmic procedure to a level which significantly exceeds that which is accessible by a unitary transformation of the thermal equilibrium state. We also show that pumped singlet order may be converted in a subsequent step to nuclear magnetization that is substantially enhanced with respect to its thermal equilibrium value, and which does therefore correspond to a reduced spin temperature.

This work exploits a very simple system: an ensemble of strongly coupled spin-1/2 pairs. It is not currently known how the phenomenon described here translates to more complex quantum systems, including those which use substrates other than nuclear spins. The fundamental limits on algorithmic cooling of the type described here, in which the relevant system eigenstates are not discrete qubit states but entangled superposition states, are not yet known. It is also unknown whether the theoretical bounds on algorithmic cooling in discrete qubit systems (*10, 11*) apply here too. We expect the current report to stimulate interest, and further development, in the quantum information community and elsewhere.

**Materials and Methods**

*Sample*. The naphthalene derivative $^{13}C_2$-**I** was synthesized by organic synthesis (*37*). The experiments used a solution of $^{13}C_2$-**I** in acetone-$d_6$ sealed into a micro-cell of length 15 mm, inserted into a standard 5 mm NMR tube and positioned about 1 cm above the bottom of the tube. This procedure avoids convection (*31*) and provides high homogeneity for both the static field and the radiofrequency field.

*NMR Spectrometer*. All experiments were performed at a magnetic field of 16.45 T using a 700MHz Bruker spectrometer.

*Cyclic permutations of spin populations*. The algorithmic cooling protocol requires a robust and high-fidelity implementation of the cyclic permutation operations for the populations of the singlet state $|1\rangle$ and the outer triplet states $|2\rangle$ and $|4\rangle$. These operations were realised by applying the pulse sequences shown in figure 6(a,b). Each permutation was implemented by applying a shaped APSOC (Adiabatic Passage Spin Order Conversion) pulse (*24–27*), followed by a composite $\pi/2$ pulse (*38*).

*APSOC pulses*. The APSOC element is an amplitude-modulated radiofrequency pulse of carefully optimized shape (*25, 26*), applied at a frequency slightly displaced from the centre of the spectral pattern. The off-resonance radiofrequency field induces a high-fidelity population exchange between the singlet state $|1\rangle$ and one of the outer triplet states $|2\rangle$ or $|4\rangle$, defined in a reference frame rotated with respect to the laboratory frame by $\pi/2$ about the rotating-frame y-axis. The shape of the APSOC pulses used here was optimized for the given spin system parameters using previously described algorithms (*25, 26*), and is well approximated by a polynomial of the form

$$\omega_{APSOC}(t) = \omega_{APSOC}^{max} \sum_{i=0}^{20} \left( \frac{t}{T_{APSOC}} \right)^i C_i \quad (19)$$

where the maximum value of the rf field amplitude, expressed as a nutation frequency, is $\omega_{APSOC}^{max} = 2\pi \times 181$ Hz, and the duration of the APSOC pulse is given by $T_{APSOC} = 0.36$ s. The APSOC pulse shape is specified by the following polynomial coefficients: {$C_0$, $C_1..C_{20}$}={ -3.58531×10$^{-3}$, 2.91521, -160.639, 6.58521×10$^3$, 145.473×10$^3$, 2.01387×10$^6$, -18.9512×10$^6$, 126.672×10$^6$, -617.009×10$^6$, 2.22025×10$^9$, -5.9162×10$^9$, 11.5203×10$^9$, -15.6681×10$^9$, 12.7611×10$^9$, -1.0036×10$^9$, -12.8512×10$^9$, 18.9193×10$^9$, -14.8478×10$^9$, 7.07406×10$^9$, -1.93568×10$^9$, 235.046×10$^6$}. The APSOC(±) pulses were applied off-resonance by ±35 Hz, relative to the centre of the NMR spectrum, taking into account the sign of the Larmor frequency (*39*), as depicted in Figure 6(a,b).

*Composite pulses*. The composite $\pi/2$ pulses were of the form $180_{\pm30}90_{\pm150}$, using the notation $\beta_\phi$ for the flip angle $\beta$ and phase $\phi$, both specified in degrees, and taking into account the sign of the Larmor frequency and the radiofrequency mixing scheme of the spectrometer (*39, 40*). The composite pulse $180_{\pm30}90_{\pm150}$ executes a rotation of spin angular momentum from the x-axis to the $\mp$z-axis, well compensated for non-idealities in the radiofrequency field amplitude (*38*).

*$T_{00}$ filter*. As its name implies, the pulse sequence element $T_{00}$ excludes signal components that do not pass through irreducible spherical tensor operators of rank 0 (*17*). In the current case of a two-spin-1/2 system the rank-0 spherical tensor operators correspond to singlet order. The sequence used here is sketched in figure 6(c), where MA indicates the magic angle, $\phi \simeq 54.74°$, and the echo durations are $\tau_\rho = 18.2$ ms. The magnetic field gradient pulses $\{G_1, G_2, G_3\}$ have amplitudes $\{0.75, 2.25, 0.51\}$ mT cm$^{-1}$ and durations $\{4.4, 2.4, 2.0\}$ ms. This implementation of the $T_{00}$ filter was taken from ref.(*29*).

**Acknowledgments:**
  **General**: M.H.L. would like to thank Soumya Singha Roy for discussions.
  **Funding:** This research was supported by EPSRC-UK (grant EP/P009980/1), the European Research Council (grant 786707-FunMagResBeacons), the Russian Foundation for Basic Research, grant 19-29-10028, the Russian Ministry of Science and Higher Education (project AAAA-A16-116121510087-5), and the European Union's Horizon 2020 research and innovation programme under the Marie Skłodowska-Curie grant agreement No 766402.
  **Author contributions:** B.A.R: implementation, experiments, theory, figures and writing; C.B: theory and writing; L.J.B. and R.C.D.B: synthetic chemistry; K. F. S., A. S. K., A. V. Y.: experiments; K.L.I: theory and writing; M.H.L.: central concept, theory, figures, simulations and writing.
  **Competing interests:** None.
  **Data and materials availability:** Data and mathematica notebook accessible through [placeholder].

**Figures and Tables**

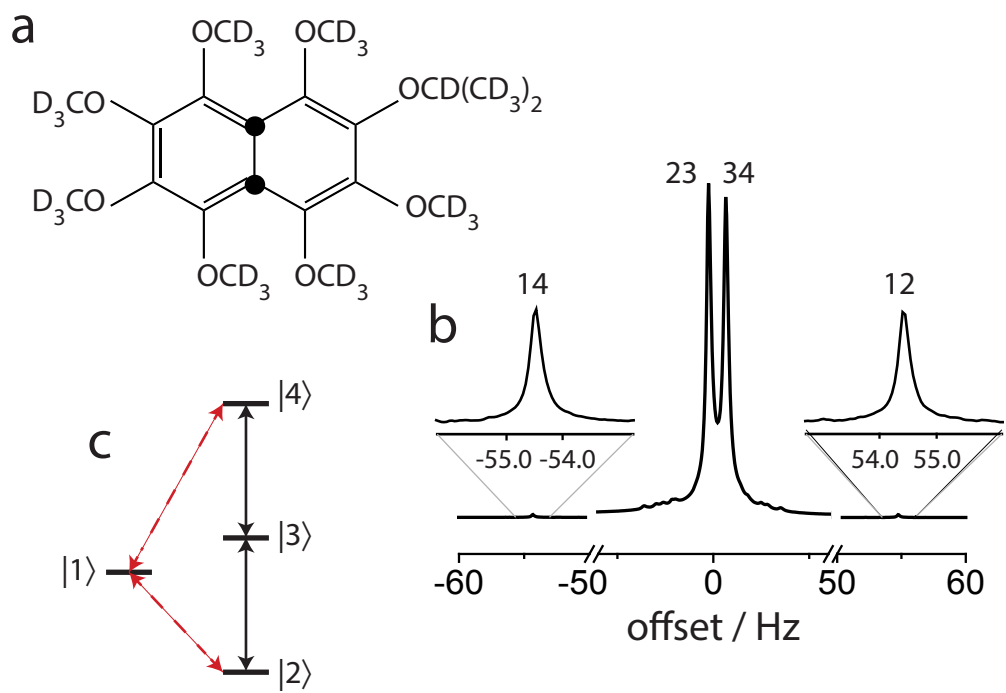

**Fig. 1. Molecular system, NMR spectrum and energy level structure.** (a) Molecular structure of the compound $^{13}C_2$-**I**, with the filled circles indicating the positions of $^{13}C$ labels. (b) $^{13}C$ NMR spectrum of a 20 mM solution of $^{13}C_2$-**I** in acetone-$d_6$ at a field of $B^0$=16.4 T. Expanded views are shown of the small outer peaks of the AB spectrum. (c) Energy level diagram of the $^{13}C_2$ pair giving the numbering of the basis states. The assignments of the observable NMR transitions to pairs of states are given in (b).

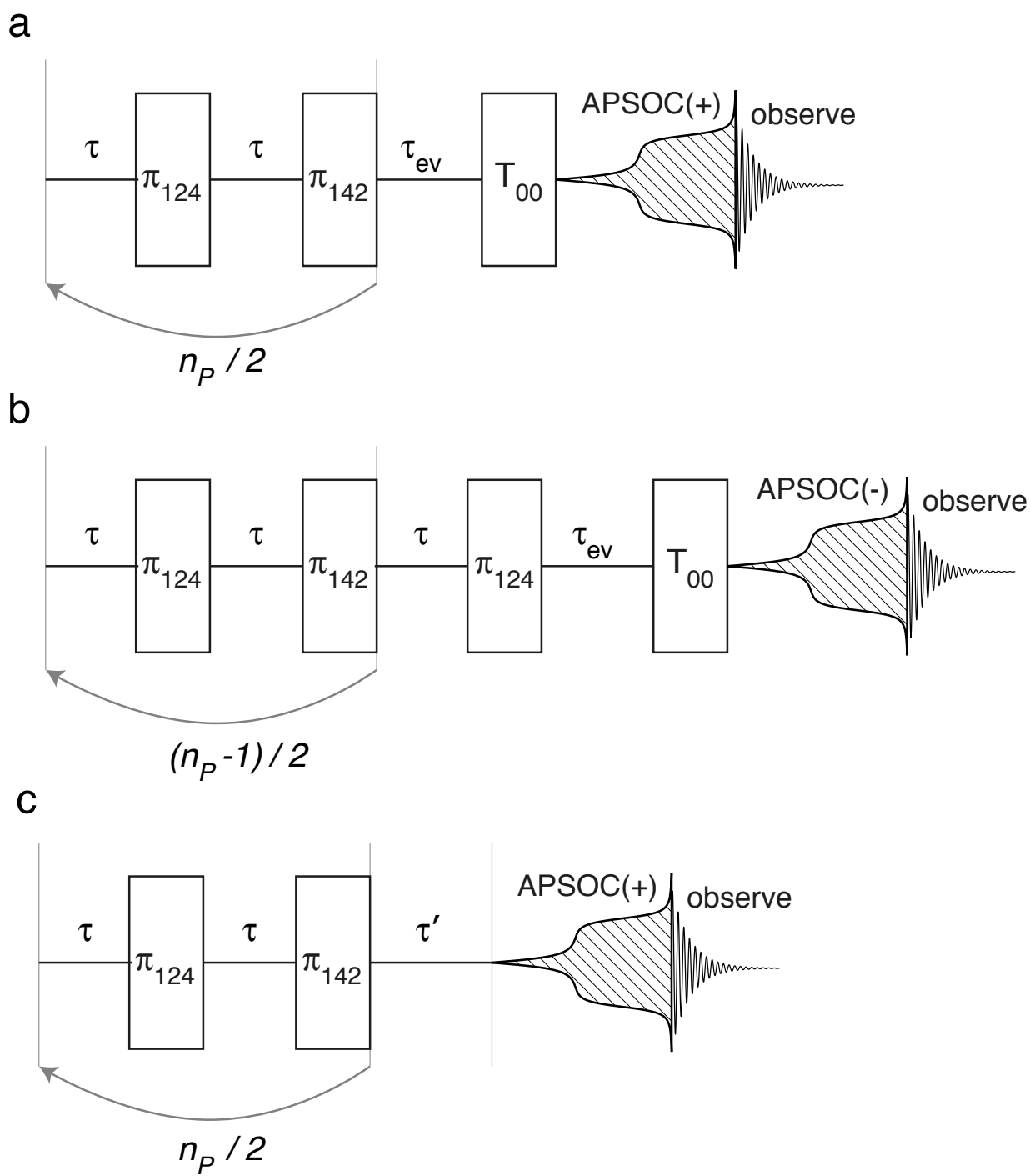

**Fig. 2. NMR Protocols.** (a) Algorithmic cooling followed by detection of enhanced singlet order for an even number of permutations $n_P$. Implementation of the permutations and APSOC pulses is shown in Fig.6. Singlet order is pumped by $n_P/2$ repetitions of the cycle $C$ (equation (11)), followed by an optional interval

$\tau_{ev}$, a $T_{00}$ filter and an APSOC(+) pulse for generation of the NMR signal. (b) Protocol for an odd number of permutations $n_P$. (c) Protocol for detecting enhanced nuclear magnetization for an even number of permutations $n_P$. A relaxation delay $\tau'$ followed by an APSOC(+) pulse is applied to the algorithmically pumped singlet order.

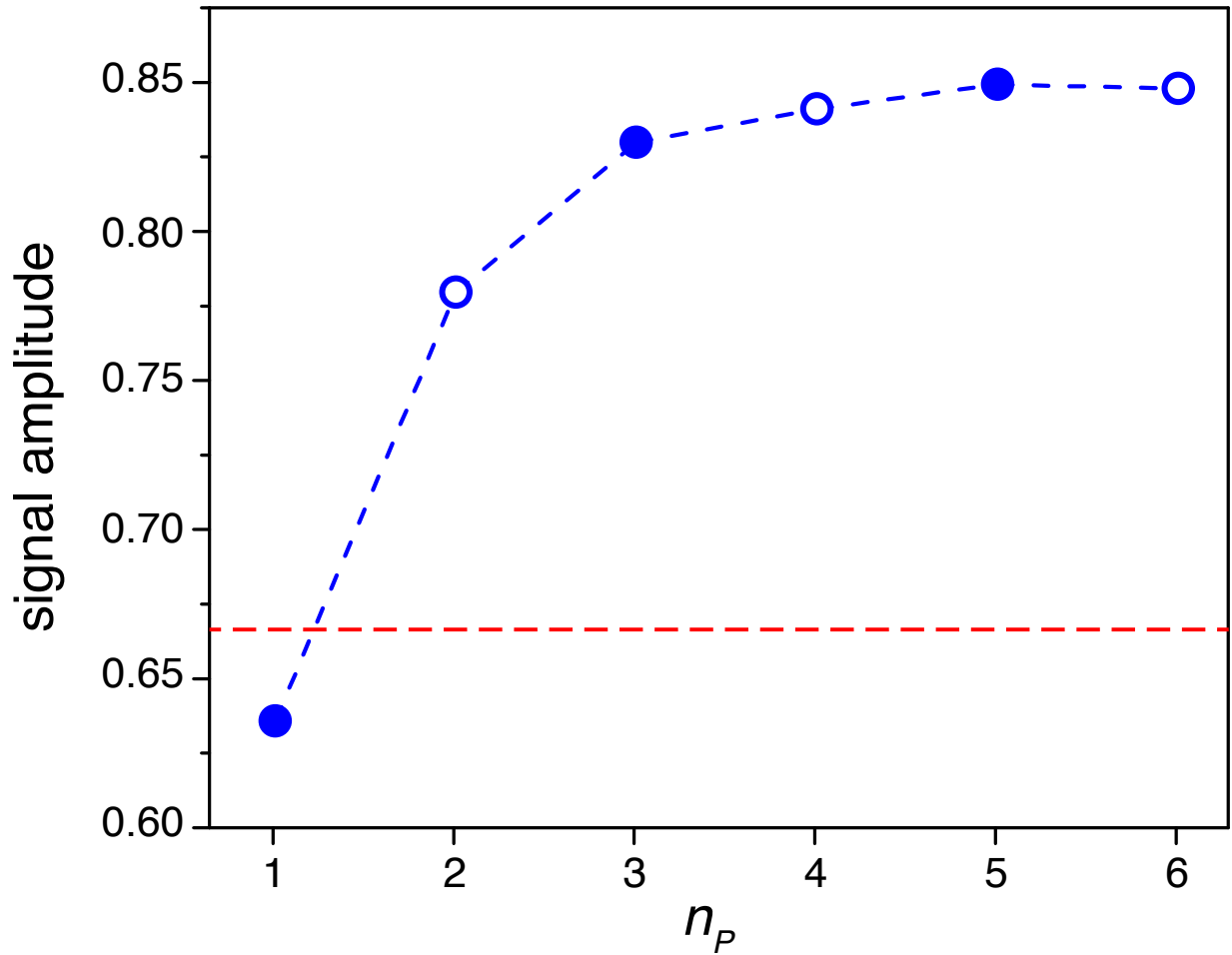

**Fig. 3. Build-up of singlet order under algorithmic cooling.** Experimental dependence of the singlet-filtered signal amplitude on the number of permutations $n_P$. Even values of $n_P$ (open symbols) and odd values of $n_P$ (filled symbols) were acquired using the protocols in Figure 2(a) and (b) respectively. The signal amplitude is normalized against the signal induced by a 90° pulse applied to a system in thermal equilibrium. The intervals for triplet thermal reset between permutations was $\tau = 28.0$ s. No evolution interval was used, $\tau_{ev} = 0$. The red dashed line indicates the theoretical bound of 2/3 for two-way magnetization-to-singlet conversion in the case of unitary spin evolution.

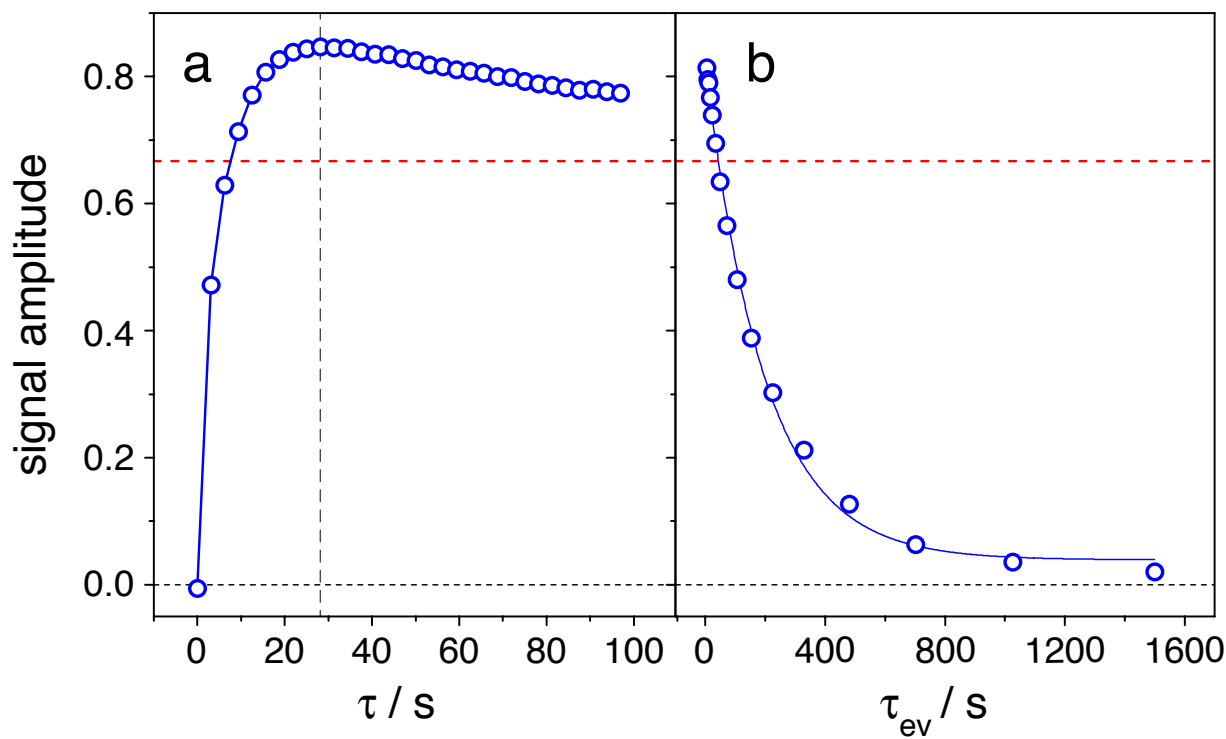

**Fig. 4. Dependence of singlet-filtered NMR signal amplitude on pulse sequence intervals.** Experimental dependence of the normalized singlet-filtered signal amplitude on the delays $\tau$ and $\tau_{ev}$. (a) Dependence on the triplet thermal reset delays $\tau$ in Figure 2(a), where $n_P = 6$ and $\tau_{ev} = 0$. The pumped singlet order is maximised for a duration $\tau = 28.0$ s (vertical dashed line). (b) Dependence on the evolution delay $\tau_{ev}$ in Figure 2(a), for $n_P = 6$ and $\tau = 28.0$ s. The decay is fitted to a monoexponential decay function with the time constant $T_S = 209 \pm 6$ s, verifying the generation of long-lived singlet order. The red dashed line indicates the theoretical bound of 2/3 for magnetization-to-singlet-to-magnetization conversion under unitary spin evolution.

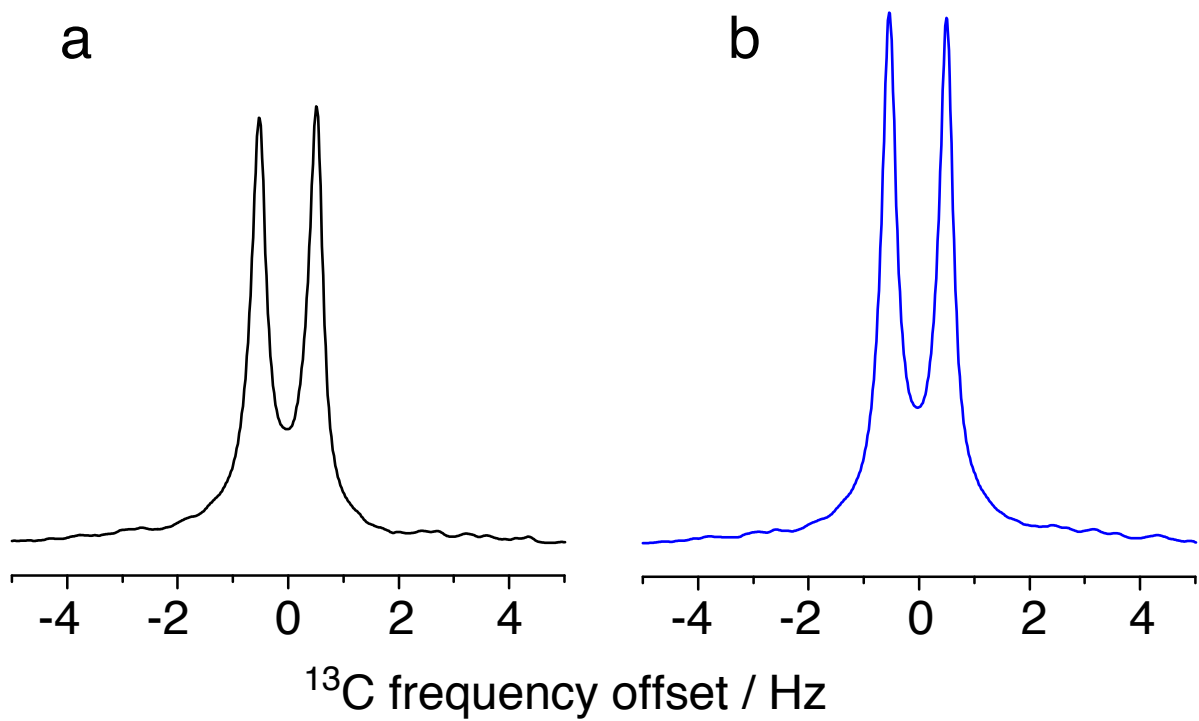

**Fig.5. NMR Signal Enhancement.** $^{13}$C NMR spectra of $^{13}$C$_2$-**I**. (a) Conventionally acquired spectrum, obtained from the free-induction decay induced by a 90° pulse applied to a system in thermal equilibrium. (b) Spectrum obtained from an algorithmically cooled spin system, obtained by the protocol in Figure 2(b), with parameters $n_P = 6$, $\tau = 28.0$ s, $\tau_{ev} = 0$, $\tau' = 18.0$ s. The signal intensity is enhanced by a factor of 1.21, with respect to thermal equilibrium.

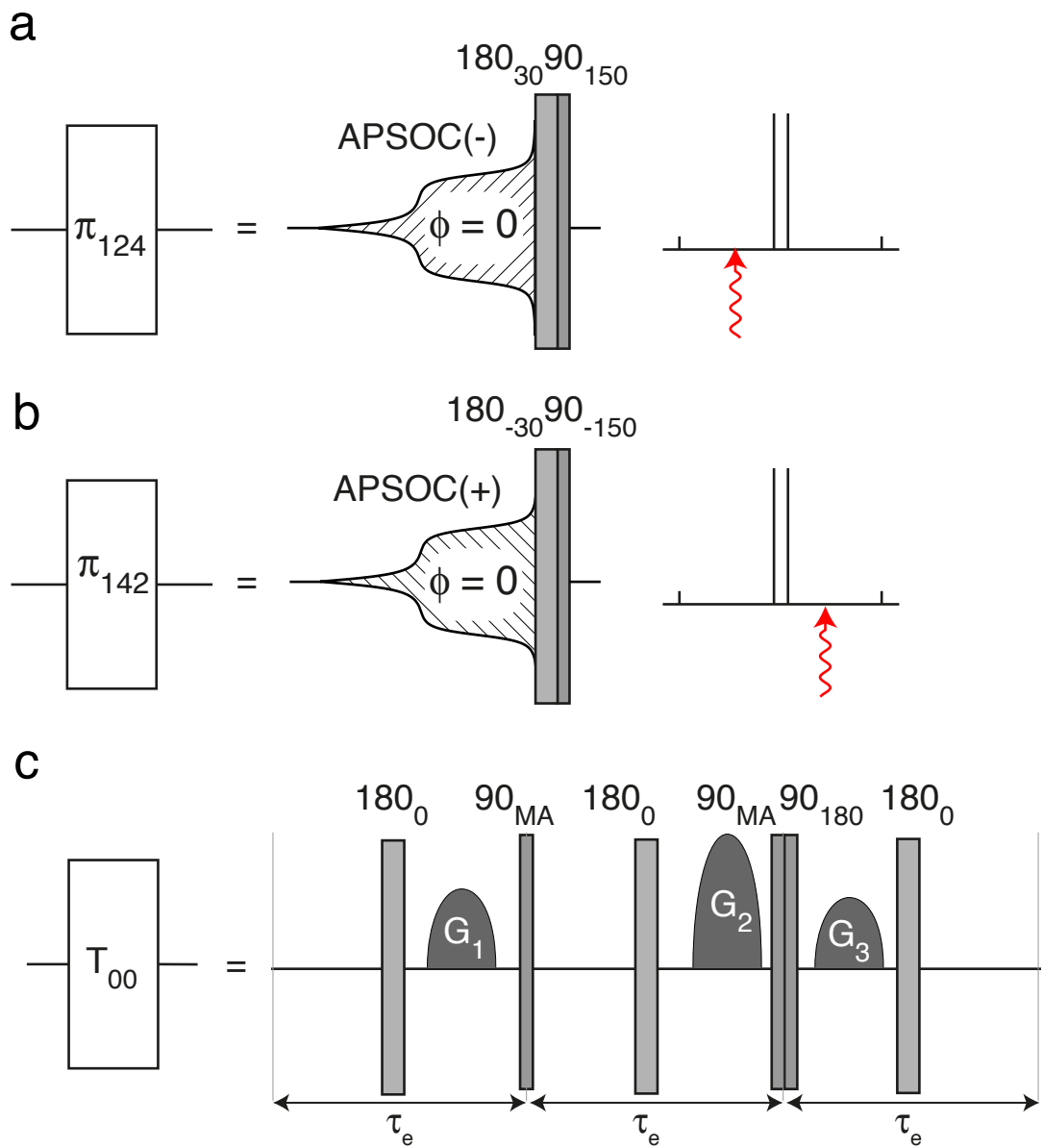

**Fig.6. Pulse Sequence Elements.** Pulse sequence elements used in this work. The APSOC (Adiabatic Passage Spin Order Conversion) pulses are specified in Methods and Materials and have phase $\phi = 0$, a duration $0.36$ s and a maximum amplitude corresponding to a $^{13}$C nutation frequency of 181 Hz. APSOC($\pm$) pulses are applied at a frequency shifted (in the algebraic sense) by $\pm 35$ Hz relative to the centre of the spectrum (red arrows in the right-hand pane). (a) The permutation element $\pi_{124}$ is implemented by applying a APSOC(-) pulse followed by a composite 90° pulse of the form $180_{30}90_{150}$. (b) The permutation element $\pi_{142}$ is implemented in a similar fashion but with the opposite frequency offset of the APSOC pulse and reversed signs of the composite pulse phases. (c) Pulse sequence for the singlet order filtration element $T_{00}$. MA denotes the magic angle (~54.74°). Field gradient pulses are denoted *{$G_1$, $G_2$, $G_3$}*.